\documentclass[11pt]{article}
\pdfoutput =1 
\PassOptionsToPackage{nosort}{cite}

\usepackage{amssymb}
\usepackage{chet}
\usepackage{mathtools}
\usepackage[scale=0.86,defaultsans]{opensans}
\usepackage{booktabs}
\usepackage{tablefootnote}

\definecolor{darkblue}{rgb}{0.1,0.1,0.7}
\hypersetup{colorlinks,
	linkcolor={darkblue},
	citecolor={darkblue},
	urlcolor={darkblue}
}

\usepackage[]{graphicx}
\usepackage{notoccite}

\usepackage{slashed}
\usepackage{dsfont}

\newcommand{\reef}[1]{(\ref{#1})}
\def\vareps{\varepsilon}
\def\eps{\epsilon}
\newcommand{\beq}{\begin{equation}} 
\newcommand{\eeq}{\end{equation}}
\def\del {\partial} 
\def\nn{\nonumber} 
\def\bZ {\mathbb{Z}} 
\def\bR {\mathbb{R}}

\def\calO {{\cal O}}

\def\ge{\geqslant}
\def\le{\leqslant}

\newcommand{\diffop}[2]{\ifthenelse{\equal{#2}{1}}{\frac{\mrm{d}}{\mrm{d} #1}}{\frac{\mrm{d}^#2}{\mrm{d} #1^#2}}}

\def\1{{\ds 1}}

\def\<{\langle}
\def\>{\rangle}

\allowdisplaybreaks

\date{}

\preprint{}

\title{A structural test for the conformal invariance of the critical 3d Ising model}

\author{Sim\~ao Meneses$^{1}$, 
	Jo\~ao Penedones$^{2}$,
	Slava Rychkov$^{3,4,5}$, \\J. M. Viana Parente Lopes$^{1}$, Pierre Yvernay$^{5}$}

\affiliation{
	$^1$ Centro de F\'{\i}sica das Universidades do Minho e Porto
	Departamento de Engenharia F\'{\i}sica, Faculdade de Engenharia and
	Departamento de F\'{\i}sica e Astronomia, Faculdade de Ci\^encias,
	Universidade do Porto, 4169-007 Porto, Portugal
	\\
	$^2$ Institute of Physics, \'Ecole Polytechnique F\'ed\'erale de Lausanne (EPFL), \\
Rte de la Sorge, BSP 728, CH-1015 Lausanne, Switzerland
	\\
	$^3$ Institut des Hautes \'Etudes Scientifiques, Bures-sur-Yvette, France
	\\
	$^4$ D\'epartement de Physique, Ecole Normale Sup\'erieure, Paris, France\\
	$^5$
	Theoretical Physics Department, CERN, Geneva, Switzerland}

\abstract{How can a renormalization group fixed point be scale invariant without being conformal? Polchinski (1988) showed that this may happen if the theory contains a virial current -- a non-conserved vector operator of dimension exactly $(d-1)$, whose divergence expresses the trace of the stress tensor. We point out that this scenario can be probed via lattice Monte Carlo simulations, using the critical 3d Ising model as an example. Our results put a lower bound $\Delta_V>5.0$ on the scaling dimension of the lowest virial current candidate $V$, well above 2 expected for the true virial current. This implies that the critical 3d Ising model has no virial current, providing a structural explanation for the conformal invariance of the model.\\[1cm]
\begin{center} \it Dedicated to the memory of Joe Polchinski (1954-2018)\\[2cm]\end{center}

\noindent v1: February 2018\\
\noindent v2: January 2019
}

\begin{document}

\maketitle

\toc

\section{Introduction}

It is believed that the critical point of the 3d ferromagnetic Ising model is conformally invariant. One strong piece of evidence is the excellent agreement between the critical exponents extracted from  experiments and Monte Carlo simulations and from the conformal bootstrap \cite{ElShowk:2012ht,El-Showk:2014dwa,Kos:2014bka,Simmons-Duffin:2015qma,Kos:2016ysd,Simmons-Duffin:2016wlq}. Conformal invariance has been also checked directly on the lattice, by verifying functional constraints that it imposes on the shape of some correlation functions  \cite{Billo:2013jda,Cosme:2015cxa}.\footnote{We would also like to point out related checks of conformal invariance in 3d self-avoiding walk \cite{Kennedy1,Kennedy2} and 3d percolation \cite{Gori:2015rta}.} In this paper we will provide another lattice test of this property, which is qualitatively different and in a sense more robust.

Any field theory coming from a local action, and in particular the 3d Ising model close to or at the critical temperature, has a local stress tensor operator $T_{\mu\nu}$ which is conserved: $\del^\mu T_{\mu\nu}=0$. The structural property of \emph{conformally invariant} local theories is that this local stress tensor operator is traceless:
\eqn
{
T_{\mu}{}^{\mu}=0\,.
}[traceless]
Our new test will probe this structural property, unlike previous lattice studies which tested its consequences.

The key question is: could the critical 3d Ising model be scale invariant (as befits any critical theory, being a fixed point of a renormalization group flow), but not fully conformally invariant? As was lucidly explained by Polchinski \cite{Polchinski:1987dy},\footnote{See also \cite{Nakayama:2013is} for a review. Concerning the 3d Ising model, see especially section 4.2 of \cite{Paulos:2015jfa}.} a theory will be scale invariant without being conformal if $T_{\mu\nu}$ is not traceless but its trace is a total divergence:
\eqn
{
T_{\mu}{}^{\mu}=\del^\nu W_\nu\,,
}[V]
where $W_\mu$ is a vector operator, called the virial current, which is (a) not conserved and (b) not itself a total derivative.\footnote{If $W_\mu$ is a total derivative, the stress tensor can be ``improved" to be traceless, so that Eq.~\traceless is satisfied for the improved $T_{\mu\nu}$.} Precisely this mechanism is responsible for scale without conformal invariance of the theory of elasticity, perhaps the simplest physically relevant example of this phenomenon \cite{Riva:2005gd}.\footnote{It should be noted that this mechanism may be realized with a quirk in gauge theories. Namely it may happen that Eq.~\V holds but that the virial current is not a gauge invariant operator (and so is not a physical local operator). For example, this is how the 3d Maxwell theory avoids conformal invariance \cite{ElShowk:2011gz}.} 

It's then natural to inquire if Eq.~\V can hold in the critical 3d Ising model, and we will show that it cannot. Our argument is based on the following simple observation: any operator $W_\mu$ which is a candidate to appear in the r.h.s.~of \V must have two additional properties. First of all, it should, just as $T_{\mu\nu}$ itself, be invariant under the internal symmetry of the model, $\bZ_2$ in the case of Ising. In addition, since $T_{\mu\nu}$ has canonical scaling dimension $d$, operator $W_\mu$ should have dimension $d-1=2$. 


For the subsequent discussion, let us define $V_{\mu}$ as the \emph{lowest} $\bZ_2$-even vector operator $V_\mu$, which is not a total derivative. If we manage to show that $\Delta_V>2$, this will imply that the model has no virial current candidates of appropriate dimension, and thus must be conformal.

 Extending the discussion from $d=3$ to the whole family of $\bZ_2$-invariant Wilson-Fisher fixed points for $2\le d\le 4$, the dimension of $V$ can be determined exactly in $d=2$ and $d=4$ (see appendix \ref{sec:dims}). Namely, we have:
 \begin{align}
 \Delta_V&=14 \quad\text{(2d Ising)},\nn\\
\Delta_V&=11\quad\text{(4d free massless scalar)}.\label{eq:exp}
 \end{align}
 It also follows from the $\eps$-expansion that the dimension of $V$ in $4-\eps$ dimensions will be $11\pm O(\eps)$.\footnote{The coefficient of the $O(\eps)$ correction term could be computed, but we don't need it.} 
 
Eqs.~\reef{eq:exp} correct some incorrect statements in the first version of this paper \cite{Meneses:2018xpuV1} and in \cite{Del,Delamotte:2018fnz,DePolsi:2018vxc}. For example, Ref.~\cite{Meneses:2018xpuV1} stated that $\Delta_V=7$ in 4d, having in mind the candidate 
\beq
V_{cand} = \phi\, \del_{\mu}\phi (\del_\nu \phi)^2\,.
\label{eq:Vcand}
\eeq
As pointed out in \cite{Delamotte:2018fnz}, this particular operator is actually total derivative, as we have the relation
 \beq
V_{cand}= \del_{\nu}[\phi^2\del_\mu\phi\del_\nu\phi]-\frac 12 \del_\mu[\phi^2(\del_\nu\phi)^2]
 \eeq
 (modulo terms vanishing by the equations of motion). However, their own dimension 7 candidate for $V$ is also incorrect, being a redundant operator (see note \ref{note:redundant}).
 
 Based on Eqs.~\reef{eq:exp}, one can expect that the dimension of $V_{\mu}$ in critical 3d Ising model should be significantly larger than 2.
 In this paper we will show, using lattice Monte Carlo simulations, that this expectation is correct.
 Namely, our analysis will imply a numerical lower bound on $\Delta_V$:
\eqn
{
	\Delta_V> 5.0\quad\text{(3d Ising)}\,.
}[Res]
In particular, this proves that $\Delta_V>2$, and shows that the 3d Ising model has no candidates for $W_\mu$. This rules out the scale without conformal invariance scenario based on \V, and thus provides a new test of conformal invariance.

The paper is structured as follows. In section \ref{sec:latticesetup}, we set up the lattice Monte Carlo simulation to measure a one-point function in a cubic lattice with peculiar boundary conditions (motivated in appendices \ref{boundary} and \ref{heuristic}).
Section \ref{sec:results} contains our numerical results 
that lead to \eqref{Res}.
We conclude with a short discussion of the implications of our result.
In appendix \ref{sec:dims}, we compute $\Delta_V$ in the 2d Ising model and in the theory of a free massless scalar in $d=4$.
In appendix \ref{sec:matching-rev}, we summarize the general procedure for matching  lattice operators with local operators of the critical field theory. This is well known among the practitioners but we do not know any good pedagogical summary in the literature.


\section{Lattice setup}
\label{sec:latticesetup}

We simulate the nearest-neighbor ferromagnetic 3d Ising model on the cubic lattice at the critical temperature.  
The Hamiltonian is 
\eqn
{
H=-\beta \sum_{\langle x y \rangle} s(x)s(y)\,,\quad s(x)=\pm 1.
}
We use the known critical temperature $\beta = \beta_c \approx  0.2216546$ \cite{Deng,Hasenbusch0}.

\subsection{Boundary conditions}

\begin{figure}[htb]
\centering
\includegraphics[width=0.5\textwidth]{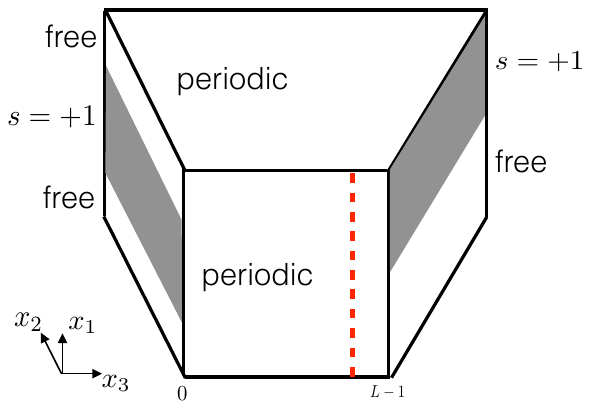}
\caption{The boundary conditions used in our simulation. The $x_3=0$ and $x_3=L-1$ faces have a combination of free (white) and fixed $s=+1$ (gray) boundary conditions. On the other faces the periodic boundary conditions are imposed. This drawing uses the Byzantine perspective only to improve visibility; the actual geometry is an $L\times L\times L$ cube with parallel sides. The red dashed line is one possible location of the integrated observable \reef{Int}. }
\label{bc}
\end{figure}

Our lattice has spatial extent $L\times L\times L$ sites. We set lattice spacing $a=1$. Due to the difficulties of measuring a rather high scaling dimension $\Delta_V$, we will only be able to go up to volumes $L=16$. We impose periodic boundary conditions in directions $x_1,x_2$, while at $x_3=0$ and $x_3= L-1$ we impose a mixture of fixed and free boundary conditions. Namely, for $x_3=0$ we impose the fixed $s=+1$ boundary condition for points with $L/4\le x_1 < 3L/4$, while at $x_3=L-1$ we do the same for points with $L/2\le x_1< L$. The rest of the boundaries at $x_3=0$ and $x_3=L-1$ has free boundary conditions (see Fig.~\ref{bc}).
The reasons for such a bizarre choice of boundary conditions 
will be explained shortly.

\subsection{Lattice operator}
We will work with the lattice operator
\eqn{
\calO^{\rm lat}_{\mu} = s(x) \nabla_\mu s(x) \sum_{\nu=1}^3 [\nabla_\nu s(x)]^2\,,
}[Vlat]
where $x$ is a lattice point and 
\eqn{
\nabla_\nu s(x) = s(x+\hat e_\nu)-s(x-\hat e_\nu)
}
is the symmetric lattice derivative in the $\nu$ direction.

Actually the precise form of the operator is unimportant, the only important thing is that $\calO^{\rm lat}_{\mu}$ is not a total lattice derivative.
See Appendix \ref{total} for a discussion and the proof of the latter fact.

\subsection{Matching of the lattice operator with critical point operators}
Close to the critical point, the lattice operator $\calO^{\rm lat}_{\mu}$ can be expanded into a basis of local operators of the critical theory with well-defined scaling dimensions (see appendix \ref{sec:matching-rev} for a review):
\eqn
{
	\calO^{\rm lat}_{\mu} = \sum_{i} c_i \calO_{i,\mu}\,,
}[Vexp]
where $\calO_i$ is the critical theory operator which has a scaling dimension $\Delta_i$, and $c_i$ are some lattice-dependent constants.
Barring accidental cancellations, any lattice measurement related to $\calO^{\rm lat}_{\mu}$ will be dominated by operators of lowest scaling dimensions appearing in the r.h.s. of \Vexp. This is because the contribution of an operator of dimension $\Delta_i$ will be suppressed by $1/R^{\Delta_i}$ where $R$ is a large distance scale (clearly  we have 
to go to large distances to explore the critical point).

Notice that operators in the r.h.s. will have to be vectors, but they don't have to be primaries.
So, the total derivative terms involving derivatives of various $\bZ_2$-even scalar operators which exist in the 3d Ising model (see Table 2 in \cite{Simmons-Duffin:2016wlq}) are expected to appear in the r.h.s. of \Vexp.
The lowest of these are $\del_\mu \vareps$ and $\del_\mu \vareps'$, where $\vareps, \vareps'$ are the lowest-dimension $\bZ_2$-even scalars, of dimension $\Delta_\vareps\approx1.41$, $\Delta_{\vareps'}\approx 3.83$. These derivative operators (especially $\del_\mu \vareps$) have rather low dimension. Below we will introduce a trick which will allow us to project them out and focus on more interesting terms.


Crucially for us, since $\calO^{\rm lat}_{\mu}$ is not a total derivative, the operator $V_\mu$ we are interested in will appear in this expansion:
\eqn
{
V^{\rm lat}_{\mu} \supset C V_\mu+\ldots\,.
}
The constant $C=O(1)$ is an unknown, non-universal, lattice quantity, and we will assume $C\ne 0$ since there is no reason to expect otherwise. The $\ldots$ include various terms which we are not interested in, and we should make sure that those terms do not mask the contribution of $V_\mu$. Some of these terms involve operators of higher scaling dimension than $V$. The presence of those terms is harmless since their effect will be subleading in the large volume limit. More annoying are the total derivative terms involving derivatives of various $\bZ_2$-even scalar operators which exist in the 3d Ising model (see Table 2 in \cite{Simmons-Duffin:2016wlq}). Some of these have a rather low dimension and would mask $V_\mu$ unless special care is taken. 
For example, we expect $\del_\mu \vareps$ to appear in the r.h.s.~of \Vexp, where $\vareps$ is the lowest-dimension $\bZ_2$-even scalar, of dimension $\Delta_\vareps\approx1.41$. 

Another class of total derivative operators which we expect to appear are $\del_\nu T'_{\mu\nu}$, 
divergences of non-conserved spin-2 $\bZ_2$-even operators. Assuming conformal invariance, the lowest such operator has dimension $\Delta_{T'}\approx 5.51$ \cite{Simmons-Duffin:2016wlq}. Divergences of higher spin operators are also expected in principle but will not play a role because of their even higher dimension. 

In our study we will be able to filter out the contributions of derivatives of scalars (like $\del_\mu \vareps$) through the following trick, rendered possible by the periodic boundary conditions. We consider the average value of the $x_1$-component of $V^{\rm lat}_{\mu}$ integrated along a periodic circle in this direction:
\eqn{
I(x_2,x_3) = \frac 1L \sum_{x_1=0}^{L-1} V^{\rm lat}_{1}(x_1,x_2,x_3)
}[Int]
Integration kills off the derivatives taken in the direction of integration. As a result this integrated observable in the continuum limit does not couple to  derivatives of scalars like $\del_\mu \vareps$. 
On the other hand divergences of spin-2 operators survive this projection, and their integral will contribute to $I$ along with the integral of $V_{\mu}$.\footnote{To kill all possible total derivatives, one could consider periodic conditions in all directions and to integrate over the whole volume. 
We do not currently have a concrete proposal implementing this idea. The main difficulty is that the one-point function of a vector operator vanishes on the 3-dimensional torus with periodic boundary conditions.} 

We will measure the one-point (1pt) function of $I$. In infinite volume vector operators would have zero 1pt functions, but in finite volume with appropriate boundary conditions they can be nonzero. In our case we will have
\eqn{
\langle I(x_2,x_3)\rangle  \equiv {\rm Obs}(x_3) =  \frac{1}{L^{\Delta_I}} f\Bigl(\frac{x_3}{L-1}\Bigr) + \ldots,
}[ff]
with no dependence on $x_2$ due to the translation invariance in that direction. The scaling of this observable with $L$ will be determined by the smaller of the two dimensions $\Delta_V,\Delta_{\del T'}=\Delta_{T'}+1$:
\beq
\Delta_I = \min(\Delta_V,\Delta_{T'}+1).
\eeq 
In this work we will only measure $\Delta_I$, but we will not be able to determine which of the two operators $V$ or $\del T'$ dominates the scaling. 

Another way to determine $\Delta_I$ would be to impose periodic boundary conditions also in the $x_3$ direction and to study finite size scaling for the 2pt function of $I$ at separation $L/2$. This observable would scale as $1/{L^{2\Delta_I}}$. We tried this strategy and found the signal completely swamped by noise, due to large $\Delta_I$. Using the 1pt function improves the signal-to-noise ratio by a factor $L^{\Delta_I}$ and will allow us to perform the measurement.

The $\ldots$ terms in \ff decay with a higher power of $L$. They originate from the higher-dimension operators contributing to $V_\mu^{\rm lat}$ as well as from corrections to scaling arising from the fact that in finite volume the theory is not exactly at the critical point but is still flowing to it in the renormalization group sense. Because of limited statistics, we will unfortunately be forced to simply neglect both of these corrections in our analysis.

The function $f(t)$, $0< t< 1$, parametrizes the observable \ff in the infinite-volume limit. This function will be measured in our simulation. To have nonzero $f(t)$, the boundary conditions at $x_3=0,L-1$ should break the flip symmetry in the $x_1$ direction:
\eqn{
x_1\to L-x_1\,,
}
under which $I$ changes sign. This is the case for our boundary conditions in Fig.~\ref{bc}. On the other hand, our boundary condition preserves the above $x_1$ flip accompanied by the $x_3$ flip:
\eqn{
x_3\to L-x_3\,,
}
and a periodic shift of the $x_1$ direction by $L/4$. As a consequence, our function $f(t)$ will be odd with respect to $t=1/2$, and in particular $f(1/2)=0$.

We have experimented with several other flip-breaking boundary conditions, and settled for the one in Fig.~\ref{bc} because it gives rise to a particularly sizable $f(t)$, thus further improving signal-to-noise. See appendix \ref{boundary} for a list of other possible boundary conditions, and appendix \ref{heuristic} for a heuristic procedure to quickly evaluate which boundary condition is expected to work best.

While it is not directly related to our computation, we would like to mention here one other instance where boundary conditions were used in lattice field theory to make a 1pt function of a tensor operator nonzero. Namely, in 4d lattice  gauge theory, the 1pt function of the off-diagonal stress tensor component $T_{0x}$ was measured imposing the ``shifted" boundary conditions, when the fields are made periodic in the spatial directions, and periodic up to a coordinate shift in the Euclidean time direction \cite{Giusti:2012yj,Giusti:2016iqr}. This boundary condition is a particular case of the gluing boundary condition discussed in appendix \ref{boundary}.


\subsection{Choice of Monte Carlo algorithm}

We perform Monte Carlo simulations using the single-spin-flip Metropolis algorithm. The choice of Monte Carlo algorithms plays a crucial role in the efficiency of the simulations. It is well
known that the Wolff algorithm \cite{Wolff} is more efficient than the Metropolis algorithm at the critical temperature due to the scaling of the computational
effort with the system size. However, even though the smaller critical slowdown exponent favors the Wolff algorithm for large systems, for small ones and for some statistical observables, the Metropolis algorithm may be more efficient. 
This is what happened in our case.

To be more concrete, the standard measure of the simulation efficiency is based
on the product of the algorithm execution time ($\tau_{CPU}$) and the integrated autocorrelation time ($\tau_{c}$).
One reason to prefer the Metropolis algorithm is that in our case it led to very small integrated autocorrelation time
of the vector operator sampling (this time scale depends on the statistical observable we
are trying to measure). 

Another important factor for this choice was the role of the boundary conditions. The use of fixed boundary conditions
requires the imposition of an acceptance probability to flip the clusters touching the boundary (see appendix \ref{boundary}).
On the other hand, if we replace the fixed b.c.~by the $\beta_{{\rm bdry}}=\infty$ conditions (see appendix \ref{heuristic}) each time a cluster
touches the boundary the full boundary will be flipped with a clear increase of $\tau_{CPU}$ and without any gain in $\tau_{c}$. These
reasons led us to opt for the Metropolis algorithm. Our tests showed that for a system size of $L=16$, the Metropolis algorithm was able to produce results with error bars comparable to the Wolff algorithm, being faster by a factor of 10.

\section{Results}
\label{sec:results}

%
%

%
%

%

We performed Monte Carlo simulations in the setup described in the previous section, with $L=8,12,16$. The nature of our boundary conditions, with the shift by $L/4$, requires to increase $L$ in steps of 4. 

Our simulations were organized as follows. To generate the next sufficiently decorrelated spin configuration we performed $N=L^3/4$ steps of the Metropolis algorithm on spins with randomly chosen positions. The measurement of the observable ${\rm Obs }(x_3)$ in \ff was then performed (averaging over $x_2$). Since our lattice operator \Vlat has range 3, we only did the measurement for $1\le x_3\le L-2$.

The total number of such decorrelated spin configurations that we generated was $2.4\times 10^{12}$ (resp.~$3.5\times 10^{13}$) for $L=12$ (resp. $L=16$). A much smaller number sufficed for $L=8$. For $N=L^3/4$ spin flips between the two measurements, the integrated autocorrelation time between the subsequent measurements of ${\rm Obs}(x_3)$ was close to 1 for every $x_3$.

Our simulations were parallelized on a cluster and took a total of about $300$ CPU-years.

\begin{table}\centering
\begin{tabular}{rrrr}
\toprule
 & \multicolumn{3}{c}{${\rm Obs}(x_3)$ in units of $10^{-6}$}\\
\cmidrule{2-4}
$x_3$ &$L=8$  & $L=12$  & $L=16$ \\
\midrule
1 & $41.9(7)$ & $9.33(17)$ & $3.11(9)$\\
2 & $12.5(7)$ & $3.12(19)$ & $1.08(9)$\\
3 & $1.7(7)$& $0.87(19)$ & $0.49(10)$\\
4 & $-3.5(7)$& $0.74(20)$ & $0.27(10)$\\
5 & $-10.7(7)$& $-0.24(20)$ & $0.12(10)$\\
6 & $-41.7(7)$& $0.16(20)$ & $-0.03(10)$\\
7 & $$ 		& $-0.39(20)$ & $0.06(10)$  \\
8 & $$		 & $-1.02(19)$ & $-0.13(10)$ \\
9 & $$ 		& $-3.18(19)$ & $-0.08(10)$ \\
10 & $$ 		& $-9.07(17)$ & $-0.07(10)$ \\
11 & $$ 		& $$ & $-0.25(10)$ \\
12 & $$		 & $$ & $-0.51(10)$ \\
13 & $$ 		& $$ & $-1.07(10)$ \\
14 & $$ 		& $$ & $-3.13(10)$ \\
\bottomrule
\end{tabular}
\caption{Results of Monte Carlo measurements with statistical errors.}
\label{MCres}
\end{table}

The numerical results of these measurements are given in table \ref{MCres}, and are shown in plots below as a function of $t=x_3/(L-1)$.\footnote{The raw data in text form can be found inside the tex file of the arxiv submission.} In these plots we show the data multiplied by $(L/12)^\Delta$ for various values of $\Delta$. According to \ff, the curves for different $L$ are supposed to collapse if $\Delta=\Delta_I$. At least this is supposed to happen for sufficiently large $L$, when contributions from the subleading terms $\ldots$ in \ff become unimportant.

\begin{figure}[htb]
\centering
\includegraphics[width=0.6\textwidth]{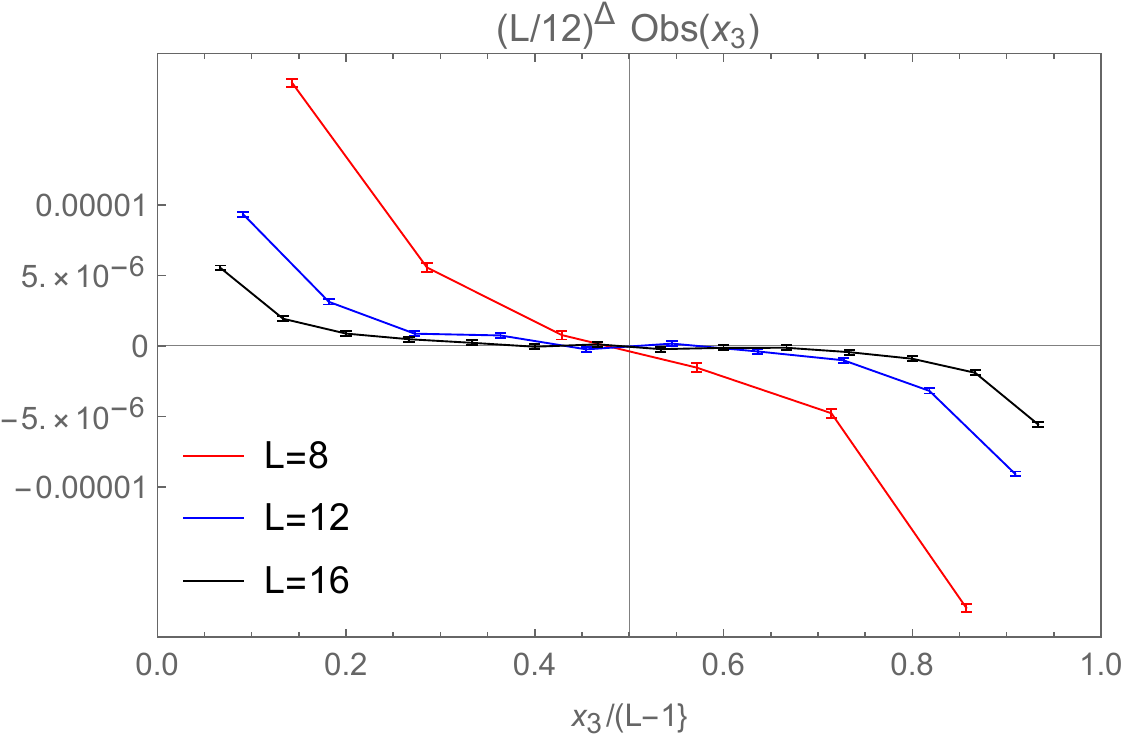}
\caption{In this plot $\Delta=2$, testing (and ruling out) the virial current existence hypothesis.}
\label{Delta2}
\end{figure}

In Fig.~\ref{Delta2} we take $\Delta=2$, the value needed for a virial current candidate.
Clearly the curves show no collapse, ruling out the existence of the virial current. 

A side remark: as mentioned in the previous section, the function $f(t)$ should be odd with respect to $t=1/2$ for our choice of the boundary conditions. This antisymmetry is indeed satisfied within error bars, as can be seen in the figures.\footnote{The way our measurement is organized, all points for the same $L$, and in particular the symmetric data points, are correlated with an unknown correlation. Thus once the measurement is finished, we cannot easily take advantage of this antisymmetry to reduce the errors by averaging over the symmetric datapoints. However, that the measured function does come out antisymmetric is a check of our procedure.}

\begin{figure}[htb]
\centering
\includegraphics[width=0.6\textwidth]{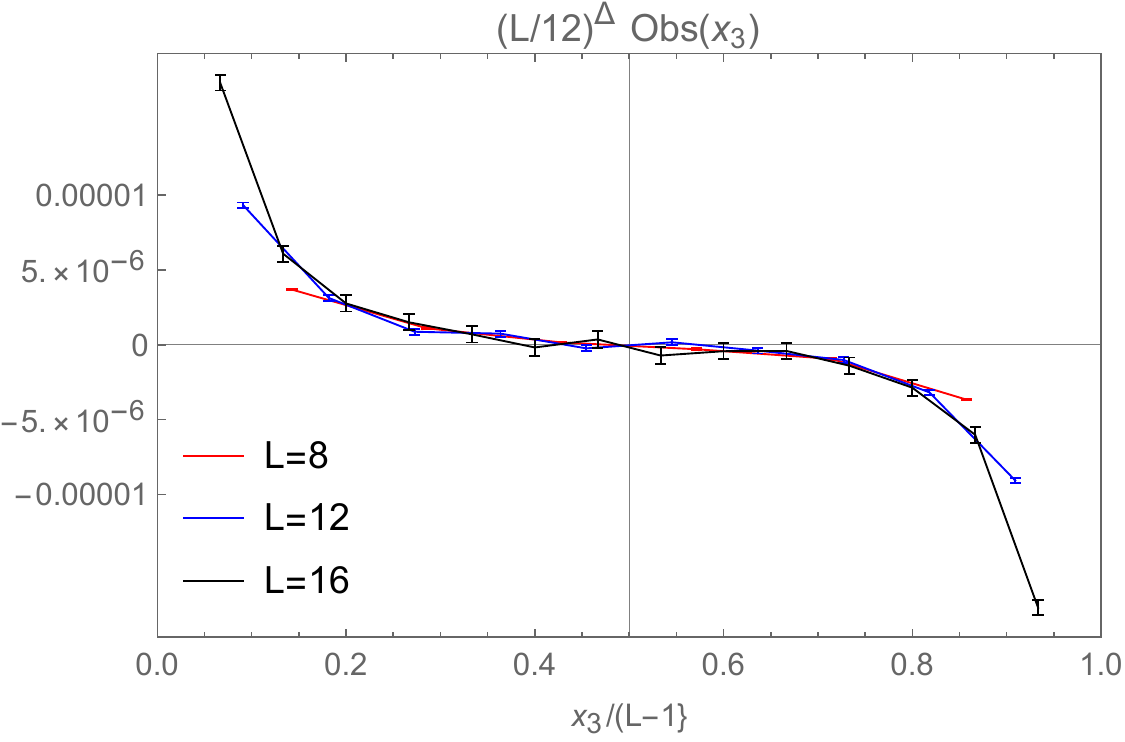}
\caption{In this plot $\Delta=6$, which is our central value for $\Delta_V$.}
\label{Delta6}
\end{figure}

In Fig.~\ref{Delta6} we show what the same plot looks like if we choose $\Delta=6$. In fact this value is our best estimate for $\Delta_I$. The curves show collapse within the error bars for $0.2\le t\le 0.8$. We consider that the $t$ values closer to the $x_3=0,L-1$ boundaries are dominated by boundary effects and exclude them from the analysis. 

\begin{figure}[htb]
\centering
\includegraphics[width=0.49\textwidth]{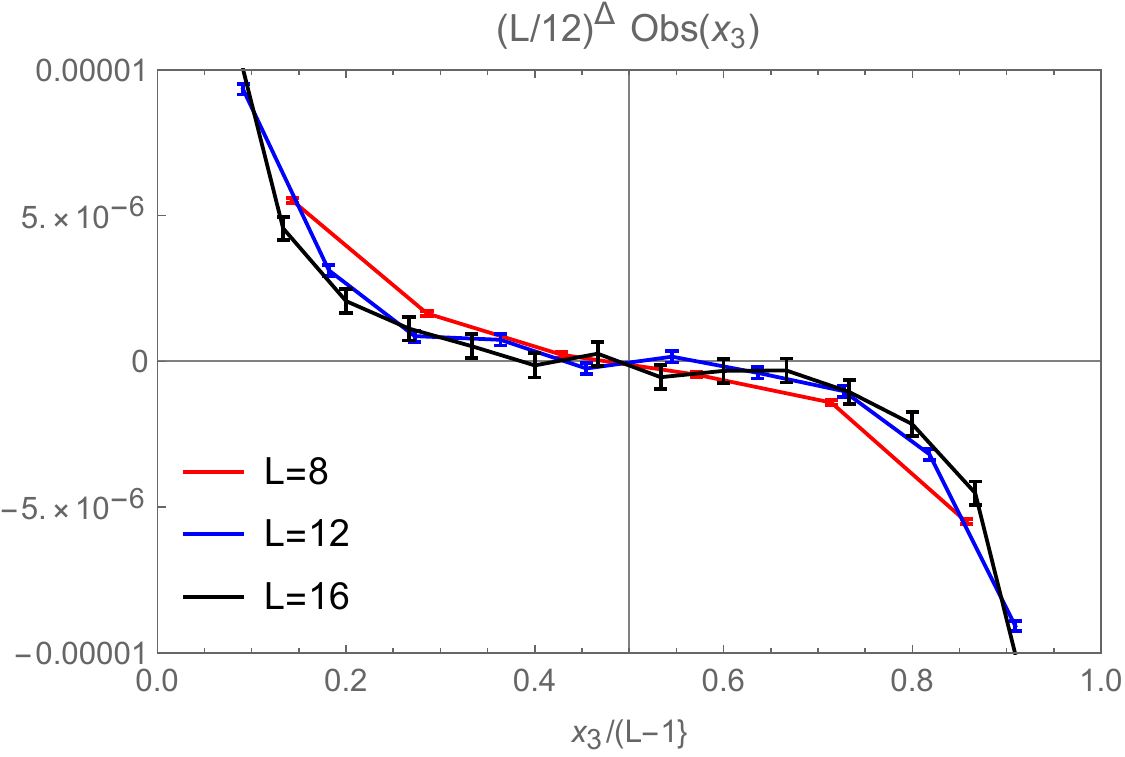}
\includegraphics[width=0.49\textwidth]{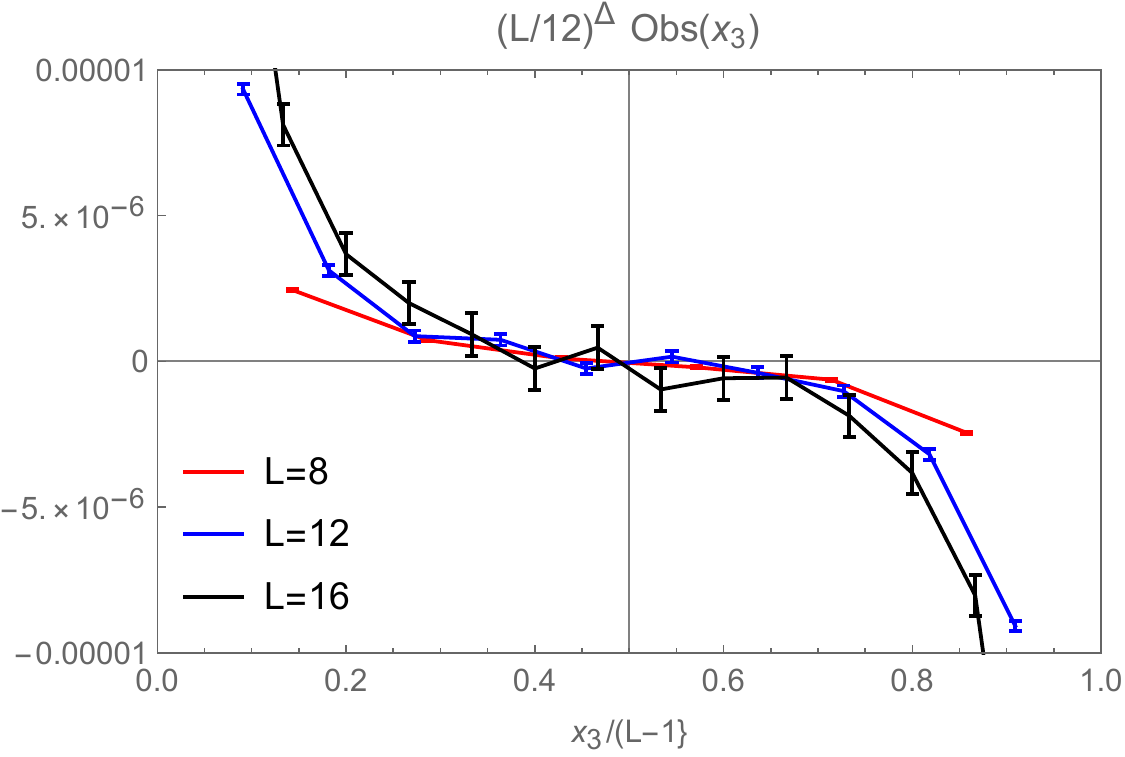}
\caption{Determining a confidence interval for $\Delta_I$. Left: $\Delta=5$. Right: $\Delta=7$.}
\label{Delta-int}
\end{figure}

To assign an error to our determination of $\Delta_I$,
we propose the following heuristic procedure. We vary $\Delta$ around 6 and see when the curves clearly deviate from the collapsing behavior in the interval $0.2\le t\le 0.8$, judging by the eye. One way to quickly perform this analysis is to use the
{\tt Manipulate} function of {\tt Mathematica}. This way we arrive at our confidence interval: 
\beq
\Delta_I=6\pm 1.
\eeq 
See Fig.~\ref{Delta-int} for what the collapse plots look like at the extreme ends of the confidence interval.\footnote{If we omit the $L=8$ datapoints from our analysis (e.g.~if one is worried that these points are still significantly affected by the subleading $\ldots$ corrections in \ff), then we get $\Delta_I=5.5\pm 1.5$ using the same procedure. We quote this number only for comparison, as we do not feel that completely discarding the $L=8$ points is justified.} 
 While the ``judging by the eye" procedure may seem subjective and ad hoc, we don't believe a much better statistical procedure can be advocated given our limited amount of data.

\begin{figure}[htb]
\centering
\includegraphics[width=0.49\textwidth]{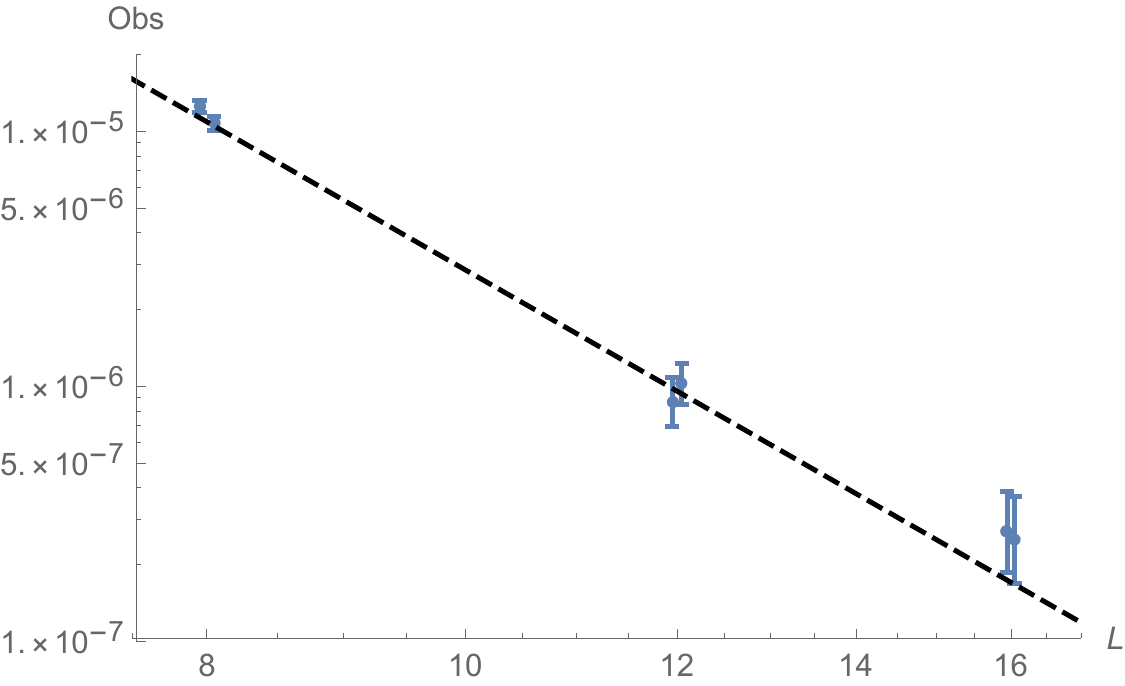}
\caption{Observable for $x_3=2$ ($L=8$), $x_3=3$ ($L=12$) and $x_3=4$ ($L=16$) and for the three mirror points (with a minus sign). The dashed line is the best fit $c/L^{\Delta}$ which gives $\Delta=6.03$ as the central value.}
\label{loglog}
\end{figure}

We have cross-checked our determination of $\Delta_I$ by focussing on the three points $x_3=2$ ($L=8$), $x_3=3$ ($L=12$) and $x_3=4$ ($L=16$), which correspond to three close values of $t=x_3/(L-1)$. Neglecting the difference in $t$, the values of the observable at these three points should scale as $const./L^{\Delta_I}$. That this is indeed roughly the case can be seen in the log-log plot in figure \ref{loglog}. Performing the fit using these three points and their mirror images under $t\to1-t$, we get the same answer $\Delta_I=6\pm 1$. 


\section{Discussion and conclusions}

One goal of this paper was to emphasize that there is a simple and robust way to check the conformal invariance of any critical lattice model, which requires the measurement of the lowest non-derivative vector operator $V$ which is a singlet under all global symmetries. This operator can play the role of the virial current, and potentially cause scale without conformal invariance, but only if its dimension is exactly $d-1$. 

In this paper we considered this strategy in the critical 3d Ising model. 
Since the dimension of $V$ appears to be large, to carry out our measurement we had to introduce several tricks increasing the efficiency of Monte Carlo simulations.
In particular, we had to consider an integrated lattice operator to decouple some uninteresting total derivative terms, and to optimize boundary conditions to maximize the (integrated) 1pt function of $V$, which was our Monte Carlo target. Further boundary condition optimization is likely possible (see appendix \ref{heuristic}) and might allow to reduce the error bars in future studies.

The main limitation of our approach to measuring $\Delta_V$ is that while it decouples total derivatives of scalars, it does not do so for divergences of spin-2 operators. As a result we measure not $\Delta_V$ but
$\Delta_I=\min(\Delta_V,\Delta_{T'}+1)$, where $T'$ is the lowest non-conserved $\bZ_2$ even spin-2. So, our result $\Delta_I=6\pm 1$ only implies a lower bound $\Delta_V\ge 5.0$ on the dimension of $V$. 
Still, the virial current value $\Delta_V=2$ is soundly ruled out by this lower bound. This confirms that the 3d Ising model is conformally invariant.

Now assuming conformal invariance, we know from the conformal bootstrap that $\Delta_{T'}\approx 5.51$ \cite{Simmons-Duffin:2016wlq}. This suggests that our measurement of $\Delta_I$ was dominated by $\Delta_{T'}+1$, while $V$ itself may be much higher. This scenario appears likely also in light of extremely high values of $\Delta_V$ in $d=2,4$ reported in the Introduction.

In this paper we have not carried out any correction-to-scaling analysis. It would be interesting to repeat the simulation in the Blume-Capel model which is in the same universality class as the Ising model but has a free parameter allowing to drastically reduce corrections to scaling \cite{Hasenbusch0}.

It would be also interesting to determine or bound the dimension of $V$ for the $O(N)$ and other models.

Finally, we would like to comment on the determination of $\Delta_V$ using the conformal bootstrap.
The numerical conformal bootstrap has determined scaling dimensions of about 100 operators of the critical 3d Ising model \cite{Simmons-Duffin:2016wlq}. The operators which have been determined appear in the operator product expansions (OPEs) of $\sigma\times\sigma$, $\vareps\times\vareps$ and $\sigma\times\vareps$, where $\sigma$ and $\vareps$ are the lowest dimension $\bZ_2$-odd and $\bZ_2$-even scalars. The OPEs $\sigma\times\sigma$ and $\vareps\times\vareps$, being OPEs of identical scalars, contain only operators of even spin. The OPE $\sigma\times\vareps$ contain only $\bZ_2$-odd operators. The operator $V$, being a $\bZ_2$-even vector, does not appear in these OPEs, and therefore it has not been so far probed by the conformal bootstrap. In the future, the OPEs $\sigma\times\sigma'$ and $\vareps\times\vareps'$, where $\sigma'$ and $\vareps'$ are the subleading $\bZ_2$-odd and $\bZ_2$-even scalars, will hopefully be included in the bootstrap analysis. These OPEs contain $V$ and can be used to determine its dimension. 

Of course, determination of $\Delta_V$ using the conformal bootstrap already presupposes that the model is conformally invariant. This has to be distinguished from the lower bound on $V$ obtained  in our paper, which is valid independently of conformal invariance, and so allowed us to test this property.

{\bf Note added.} In the first arXiv version of this paper \cite{Meneses:2018xpuV1} the reader will find an appendix criticizing the argument in \cite{Del} for  conformal invariance of the critical 3d Ising model. We consider the objections raised there still valid, and the rebuttal \cite{Delamotte:2018fnz} unsatisfactory.
However, we  removed the appendix to keep the focus on the positive results obtained in our own work. 

\section*{Acknowledgements}
We would like to thank Leonardo Giusti and Agostino Patella for the useful discussions, and Matthijs Hogervorst, Marco Meineri, and Agostino Patella for comments and suggestions for the draft. 
PY is grateful to the CERN Theoretical Physics Department for hospitality. 
JP, SR and PY were supported by the National Centre of Competence in Research SwissMAP funded by the Swiss National Science Foundation. 
This research received funding from the Simons
Foundation grants JP:\#488649 and SR:\#488655 (Simons
collaboration on the Non-perturbative bootstrap). 
SR is supported by Mitsubishi Heavy Industries as an ENS-MHI Chair holder. 
Calculations of this paper were performed at the CERN Theory cluster. We are grateful to the authors of ALPS \cite{Alps1,Alps2} whose library was very useful for setting up preliminary simulations.

\appendix
\section{Theoretical expectations for the dimension of $V$}
\label{sec:dims}

In this appendix, we determine the  lowest dimension of a vector primary operator at the Wilson-Fisher fixed point in spacetime dimension $d=2$ and $d=4$. These exactly solvable cases provide an indication for what to expect in $d=3$. 

\subsection{Four dimensions}

The Wilson-Fisher fixed point in $d=4$ describes a free massless scalar field $\phi$ satisfying the equation of motion $\partial^2 \phi=0$.
The operator content of this free CFT 
can be encoded in the partition function 
\beq
Z(q,x,y)= \sum_\mathcal{O} q^{\Delta_\mathcal{O}} x^{2j_\mathcal{O}} y^{2\bar{j}_\mathcal{O}}\,,
\eeq
where the sum runs over all local operators. The quantum numbers $(\Delta, j,\bar{j})$ are the eigenvalues of the dilatation generator $D$ and two commuting rotation generators $J_3$ and $\bar{J}_3$. The latter correspond to the decomposition $SO(4)=SU(2)\times SU(2)$ of the rotation group.
The partition function can be easily computed using the Fock space structure \cite{Barabanschikov:2005ri, Liendo:2017wsn}. We start by introducing the partition function $z_\phi$ of local operators with a single field $\phi$ and arbitrary number of derivatives,
\beq
z_\phi (q,x,y)= \chi_{1,0,0} (q,x,y) -\chi_{3,0,0} (q,x,y)
\eeq
where 
\beq
\chi_{\Delta,\ell,\bar{\ell}} (q,x,y) = 
\frac{q^\Delta}{(1-q x y)(1-q y /x)(1-qx/y)(1-q/(xy))} \sum_{j=-\ell}^{\ell}x^{2j}
\sum_{\bar{j}=-\bar{\ell}}^{\bar{\ell}}x^{2\bar{j}}
\eeq
is the long character of a conformal multiplet with primary of dimension $\Delta$ and spin $(\ell,\bar{\ell})$.
The full partition function can then be written as
\beq
Z(q,x,y)= \exp \left[ \sum_{k=1}^\infty \frac{1}{k} 
z_\phi \left(q^k,x^k,y^k\right) \right]\,.
\eeq
Moreover, the partition function restricted to $\mathbb{Z}_2$ even/odd operators is given by
\beq
Z_{\pm}(q,x,y)= \frac{1}{2} \exp \left[ \sum_{k=1}^\infty \frac{1}{k} 
z_\phi \left(q^k,x^k,y^k\right) \right]\pm
\frac{1}{2} \exp \left[ \sum_{k=1}^\infty \frac{(-1)^k}{k} 
z_\phi \left(q^k,x^k,y^k\right) \right]
\,.
\eeq
We are interested in the character decomposition of the $\mathbb{Z}_2$ even partition function. Expanding the given expression and matching the powers of $q$ and dependence on $x,y$ order by order, we arrive at the following expression:
\begin{align}
Z_{+} &= 1+ \sum_{n=1}^4 \chi^{short}_{2+2n,n,n}+
\chi _{2,0,0}+\chi _{4,0,0}+\chi
   _{6,0,0}+\chi _{6,1,1}+\chi
   _{7,\frac{3}{2},\frac{3}{2}}\\ \nonumber
   &+2
   \chi _{8,0,0}+\chi _{8,0,2}+2
   \chi _{8,1,1}+\chi _{8,2,0}+2
   \chi _{8,2,2}\\ \nonumber&+\chi
   _{9,\frac{1}{2},\frac{3}{2}}+\chi
   _{9,\frac{1}{2},\frac{5}{2}}+\chi
   _{9,\frac{3}{2},\frac{1}{2}}+\chi
   _{9,\frac{3}{2},\frac{3}{2}}+\chi
   _{9,\frac{3}{2},\frac{5}{2}}+\chi
   _{9,\frac{5}{2},\frac{1}{2}}+\chi
   _{9,\frac{5}{2},\frac{3}{2}}+\chi
   _{9,\frac{5}{2},\frac{5}{2}}\\ \nonumber&+3
   \chi _{10,0,0}+\chi
   _{10,0,2}+4 \chi
   _{10,1,1}+\chi _{10,1,2}+2
   \chi _{10,1,3}+\chi
   _{10,2,0}+\chi _{10,2,1}\\&\nonumber+4
   \chi _{10,2,2}+\chi
   _{10,2,3}+2 \chi
   _{10,3,1}+\chi _{10,3,2}+3
   \chi _{10,3,3}\\&\nonumber+ \textcolor{blue}{\chi
   _{11,\frac{1}{2},\frac{1}{2}}}+
   2 \chi
   _{11,\frac{1}{2},\frac{3}{2}}+
   2 \chi
   _{11,\frac{1}{2},\frac{5}{2}}+
   \chi
   _{11,\frac{1}{2},\frac{7}{2}}+
   2 \chi
   _{11,\frac{3}{2},\frac{1}{2}}+
   4 \chi
   _{11,\frac{3}{2},\frac{3}{2}}+
   3 \chi
   _{11,\frac{3}{2},\frac{5}{2}}+
   2 \chi
   _{11,\frac{3}{2},\frac{7}{2}}\\&\nonumber+
   2 \chi
   _{11,\frac{5}{2},\frac{1}{2}}+
   3 \chi
   _{11,\frac{5}{2},\frac{3}{2}}+
   3 \chi
   _{11,\frac{5}{2},\frac{5}{2}}+
   2 \chi
   _{11,\frac{5}{2},\frac{7}{2}}+
   \chi
   _{11,\frac{7}{2},\frac{1}{2}}+
   2 \chi
   _{11,\frac{7}{2},\frac{3}{2}}+
   2 \chi
   _{11,\frac{7}{2},\frac{5}{2}}+
   2 \chi
   _{11,\frac{7}{2},\frac{7}{2}} \\& +O(q^{12}), \nonumber
\end{align}
where
\beq
\chi^{short}_{2+2n,n,n} =\chi_{2+2n,n,n}-\chi_{3+2n,n-\frac{1}{2},n-\frac{1}{2}}
\eeq
is the character associated with a conserved current of spin $2n$.
This shows that the vector primary with lowest scaling dimension has $\Delta=11$ (blue character).

As a consistency check, we have determined $\Delta_V=11$ using an alternative method. We performed the conformal block decomposition of the four-point function \footnote{We normalized the operators $\phi^2$ and $\phi^4$ to have unit two-point function.}
\begin{multline}
\langle \phi^2(x_1) \phi^4(x_2)\phi^2(x_3)\phi^4(x_4) \rangle 
=
\frac{1}{x_{13}^4 x_{24}^8}+\frac{6}{x_{12}^4 x_{34}^4 x_{24}^4}
+\frac{6}{x_{14}^4 x_{23}^4 x_{24}^4}\\+
\frac{8}{x_{13}^2 x_{12}^2 x_{34}^2 x_{24}^6}
+\frac{8}{x_{13}^2 x_{14}^2 x_{23}^2 x_{24}^6}
+\frac{24}{x_{14}^4 x_{12}^2 x_{34}^2 x_{23}^4 x_{24}^4}\,.
\end{multline}
In the (12) channel, the conformal block decomposition reads
\footnote{We use the standard conformal block as defined in \cite{Dolan:2000ut, Costa:2011dw}.}
\begin{align}
&6 G_{2,0}+32 G_{4,0}+15
   G_{6,0}+\frac{96 }{5} G_{6,2}+8
   G_{7,3}+\frac{128
   }{7}G_{8,2}+\frac{384
  }{35} G_{8,4}+\frac{16
   }{5}G_{9,3}+\frac{64
   }{11}G_{9,5}\\ &+\frac{2
   }{5}G_{10,0}+\frac{12
  }{7} G_{10,2}+\frac{464
   }{33}G_{10,4}+\frac{15872
   }{3003}G_{10,6}+\frac{8
   }{25}\textcolor{blue}{G_{11,1}}+G_{11,3}+\frac{3
   84 }{91}G_{11,5}+\frac{192
   }{65}G_{11,7} +\dots \nonumber
\end{align}
where $G_{\Delta,s}$ stands for the conformal block of dimension $\Delta$ and spin $s$ (corresponding to the SO(4) irreducible representation $(\frac{s}{2}, \frac{s}{2})$). Again we find the first vector primary at dimension 11.

  One can also see that the vector primary operator we identified is parity-even. This follows immediately because parity odd vector primary operators cannot appear in the OPE of two scalars (like $\phi^2$ and $\phi^4$) in a parity symmetric theory.
 In addition, it is easy  to see that the vector operator contains 6 fields $\phi$ and 5 derivatives. \footnote{The $\phi$ content of each primary can be obtained by studying the partition function 
 \beq
Z(r,q,x,y)= \exp \left[ \sum_{k=1}^\infty \frac{r^k}{k} 
z_\phi \left(q^k,x^k,y^k\right) \right]\,,
\eeq
where $r$ is a fugacity for the number of $\phi$'s in each local operator.}
We also studied the conformal character decomposition of the free massless scalar in $d=3$. The lightest vector primary still contains 6 fields $\phi$ and 5 derivatives, which leads to $\Delta_V=8$ in $d=3$.

The conclusion that the lowest $\bZ_2$ even vector primary has dimension 11 was reached independently by Marco Meineri \cite{Marco}. He used a different approach, which also provides an explicit expression for this primary in terms of $\phi$ and its derivatives. In $d=4-\epsilon$, this vector primary operator will get an $O(\epsilon)$ anomalous dimension, computable starting from an explicit expression in \cite{Marco}; this will not be done here.

One potential worry could be the recombination of this multiplet   with a short multiplet when $\epsilon>0$. However, it is well known (see e.g.~\cite{Rychkov:2015naa} for a discussion) that the only multiplets that recombine are
the multiplet of $\phi$ with the one of $\phi^3$ and the multiplets $\chi^{short}_{2+2n,n,n}$ (conserved currents of spin $2n$) with $\chi_{3+2n,n-\frac{1}{2},n-\frac{1}{2}}$ for $n=2,3,\dots$. So the vector primary of dimension 11 will survive as a vector primary of dimension $11+O(\eps)$ in $4-\eps$ dimensions.

Notice that in all the above discussion we set $\del^2\phi=0$ in 4d, eliminating operators involving the letter ``$\del^2\phi$" from consideration. When we go to $(4-\eps)$ dimensions, we will have the  equation of motion $\del^2\phi\propto \phi^3$. So when classifying the local operators in $(4-\eps)$ dimensions, it would be double counting to consider operators involving $\del^2\phi$. Operators proportional to the equations of motion are known as ``redundant operators" \cite{Wegner-redundant,DombGreenVol6}. While such ``operators" are useful in formal treatments of renormalized perturbation theory \cite{Collins}, they have correlation functions which are zero except at coincident points, and their dimensions do not correspond to critical exponents measurable e.g.~in lattice simulations. So redundant operators do not count as local operators of the critical theory.\footnote{\label{note:redundant}As a side remark, we note that the ``exact critical exponents'' discussed in Ref.~\cite{DePolsi:2018vxc} correspond in fact to redundant operators, making the discussion of that paper of little relevance to the physics of the Ising critical point.}

%

\subsubsection{Evanescent operators}

Here we will discuss, and exclude, the possibility, that the lowest primary vector in $4-\eps$ dimension is not the vector primary of dimension $11+O(\eps)$ discussed above, but a still lower vector primary which is an evanescent operator. Recall that the evanescent operators are those which do not exist in $d=4$ but only in $d=4-\eps$, see \cite{Hogervorst:2015akt} for a discussion. The evanescent operators arise because of antisymmetrization of indices, which kills an operator in $d=4$. Thus, they have to involve a contraction with
\beq
\delta^{\mu_1[\nu_1}\delta^{|\mu_2|\nu_2}\ldots \delta^{|\mu_5|\nu_5]}
\eeq
which in integer dimensions becomes
\beq
\eps^{\mu_1\mu_2\ldots \mu_5} \eps^{\nu_1\nu_2\ldots \nu_5}\,.
\eeq
Any operator involving this contraction will vanish identically in $d=4$, because the index $\mu$ runs only over 4 values.

The lowest vector operator which vanishes in $d=4$ but not in $d=4-\eps$ is \cite{Hogervorst:2015akt}
\beq
\delta^{\mu_1[\nu_1}\delta^{|\mu_2|\nu_2}\ldots \delta^{|\mu_5|\nu_5]}\del_{\mu_1}\phi\, \del_{\mu_2}\del_{\nu_2}\phi\ldots 
\del_{\mu_5}\del_{\nu_5}\phi,
\eeq
of dimension $14+O(\eps)$. This operator is not a primary \cite{Hogervorst:2015akt}, so the lowest evanescent vector primary is still somewhere higher. We conclude that the evanescent operators cannot compete with the $11+O(\eps)$ primary that we found above.

\subsection{Two dimensions}

Here we discuss spectrum of the 2d Ising model in the $\bZ_2$-even sector.
The Ising model contains 2 $\bZ_2$-even Virasoro primaries, $\mathds{1}$ with $h=\bar h=0$ and $\eps$ with $h=\bar h=\frac{1}{2}$.
Their Virasoro characters are given by 
\beq
\chi_\mathds{1}(q,\bar q)=\chi_0(q)\chi_0(\bar q), \qquad
\chi_{\eps}(q,\bar q)=\chi_{\frac{1}{2}}(q)\chi_{\frac{1}{2}}(\bar q)\,.
\eeq
The characters $\chi_0$ and $\chi_{\frac{1}{2}}$ 
are given by \cite{ChristeHenkel}
\begin{gather}
\chi_0(q)=1 + q^2 + q^3 + 2 q^4 + 2 q^5 + 3 q^6 + 3 q^7 + 5 q^8 + 5 q^9 + 
 7 q^{10} + 8 q^{11} + 11 q^{12}+\ldots\\
 \chi_{\frac{1}{2}}(q)= q^{\frac{1}{2}} (1 + q + q^2 + q^3 + 2 q^4 + 2 q^5 + 3 q^6 + 4 q^7 + 5 q^8 + 
   6 q^9 + 8 q^{10} + 9 q^{11} + 12 q^{12}+\ldots)
   \nonumber
\end{gather}
These Virasoro characteres can de decomposed into characters 
\beq
X_h(q)=\frac{q^h}{1-q}\,,
\eeq
of the global conformal algebra.
%
This gives 
\begin{gather}
\chi_0= 1+X_2+X_4+X_6+2X_8+\ldots\\
\chi_{\frac{1}{2}}=X_{\frac{1}{2}}+X_{\frac{9}{2}}+X_{\frac{13}{2}}+X_{\frac{15}{2}}+X_{\frac{17}{2}}+\ldots
\end{gather}

The first vector quasiprimary is obtained by combining $X_h$ with $X_{\bar h}$ with $h-\bar h=1$. We see that the minimal choice is $h=\frac{15}{2}, \bar h=\frac{13}{2}$, corresponding to the scaling dimension $\Delta=h+\bar h=14$.
It is also interesting to find a dimension of the first non-conserved spin-2 quasiprimary, for which we need $h-\bar h=2$. This is possible for $h=4,\bar h=2$, which gives $\Delta=6$.
 
The vector quasiprimaries can also be found by studying the (global) conformal block decomposition of a four-point function involving two different scalar operators. In 2d Ising, the simplest choice is $\epsilon$ (with $\Delta=1$) and $T\bar{T}$ (with $\Delta=4$). Such correlation functions can be easily computed using the conformal Ward identities.
In particular, we obtained
\beq
A(z,\bar{z})=\lim_{w \to \infty} |w|^8 \langle
\epsilon(0,0)\, T\bar{T}(z,\bar{z})\, \epsilon(1,1) \,
T\bar{T}(w,\bar{w}) \rangle
=\frac{1}{16}\left| 1+\frac{(1-2z)^2}{z^2(1-z)^2} \right|^2\,.
\eeq
The conformal block expansion in the $z,\bar{z} \to 0$ channel is given by
\begin{align}
A=&\frac{1}{16}G_{1,0}+G_{5,4}+\frac{
   4 }{5}G_{7,6}+\frac{32
   }{429}G_{8,7}+\frac{1}{16}G_{9,0}+\frac{16
   }{35}G_{9,8}+\frac{16
   }{221}G_{10,9}+\frac{1}
   {20}G_{11,2}+\frac{640
   }{2907}G_{11,10}\nonumber \\
   &+\frac{2
   }{429}G_{12,3}+\frac{512
 }{11305}  G_{12,11}+\frac{1}{400}G_{13,
   0}+\frac{1}{35}G_{13,4}+\frac{3200
   }{33649}G_{13,12}\\&+\frac{1}{4290}\textcolor{blue}{G_{14,
   1}}+\frac{1}{221}G_{14,5}
   +\frac{512
   }{22287}G_{14,13}+\dots \nonumber
\end{align}
in terms of conformal blocks \cite{Dolan:2000ut}
\beq
G_{\Delta,s}(z,\bar{z}) = \frac{k_{\Delta+s} (z) k_{\Delta-s} (\bar{z}) +k_{\Delta-s} (z) k_{\Delta+s} (\bar{z}) }{2^s\left(1+\delta_{s,0}\right)}\,,\qquad
k_\beta(z)=(-z)^\frac{\beta-9}{2} \ _2F_1\left(\frac{\beta+3}{2},\frac{\beta-3}{2},\beta,z\right)\,
\nonumber
\eeq
(shifts in the familiar exponents w.r.t.~$\beta/2$ due to unequal dimensions of external scalars).
This confirms that $\Delta_V=14$ in the 2d Ising CFT.

\section{Why $\calO^{\rm lat}_{\mu}$ is not a total lattice derivative}
\label{total}

By definition, a lattice operator $A$ is a total lattice derivative (TLD) if it can be written as the difference of a lattice operator and its translation by some fixed lattice distance, or more generally a linear combination theoreof:
\beq
A(x)= \sum_i [B_i(x)-B_i(x+y_i)]
\eeq
where $B_i$'s are some lattice operators, and $y_i$ are some lattice vectors. A multi-component operator, like $\calO^{\rm lat}_{\mu}$, is a TLD, if each of its components is a TLD (where $B_i$ and $y_i$ will depend on the component).

An obvious example of a TLD operator is $\nabla_\nu s(x)$. A less obvious example is $s(x) \nabla_\nu s(x)$, since it can be written as
\beq
s(x) \nabla_\nu s(x) = s(x)s(x+\hat e_\nu) - s(x-\hat e_\nu) s(x) = B_\nu (x)-B_\nu (x-\hat e_\nu),
\eeq
where $B_\nu(x)= s(x)s(x+\hat e_\nu)$. 

Consider now our operator $\calO^{\rm lat}_{\mu}$, focussing for definiteness on its component $\mu=1$. Using the fact that $s(x)^2=1$ for the Ising spins, it's easy to see that
\beq
\calO^{\rm lat}_1(x)=-2 A^{(1)}(x)+A^{(2)}(x)
\eeq
where 
\beq
A^{(1)}(x) = s(x)[s(x+\hat e_1)-s(x-\hat e_1)][s(x+\hat e_2)s(x-\hat e_2) + s(x+\hat e_3)s(x-\hat e_3)]
\eeq
and $A^{(2)}(x)= 8 s(x)\nabla_1 s(x)$ is a TLD operator.

We claim that $A^{(1)}$ is NOT a TLD operator. To prove this, consider the following configuration of spins:
\beq
\begin{array}{cccccccc}
	1 & 1 & 1 & 1 & 1 & 1 & 1 & 1 \\
	1 & 1 & 1 & 1 & 1 & 1 & 1 & 1 \\
	1 & 1 & 1 & 1 & 1 & 1 & 1 & 1 \\
	1 & 1 & 1 & 1 & -1 & 1 & 1 & 1 \\
	1 & 1 & 1 & -1 & -1 & 1 & 1 & 1 \\
	1 & 1 & 1 & 1 & 1 & 1 & 1 & 1 \\
	1 & 1 & 1 & 1 & 1 & 1 & 1 & 1 \\
	1 & 1 & 1 & 1 & 1 & 1 & 1 & 1 \\
\end{array}
\eeq
where we show only a slice of the 3D configuration in the $(x_1,x_2)$ plane. It is assumed that the spins are constant in $x_3$ direction, and that the lattice is periodic in all directions (we consider periodic lattice just for this proof, Monte Carlo simulations are done with different boundary conditions). Computing $A^{(1)}$ operator in this configuration, we find:
\beq
\begin{array}{cccccccc}
	0 & 0 & 0 & 0 & 0 & 0 & 0 & 0 \\
	0 & 0 & 0 & 0 & 0 & 0 & 0 & 0 \\
	0 & 0 & 0 & 0 & 0 & 0 & 0 & 0 \\
	0 & 0 & 0 & 0 & 0 & 4 & 0 & 0 \\
	0 & 0 & -4 & 4 & 0 & 4 & 0 & 0 \\
	0 & 0 & 0 & 0 & 0 & 0 & 0 & 0 \\
	0 & 0 & 0 & 0 & 0 & 0 & 0 & 0 \\
	0 & 0 & 0 & 0 & 0 & 0 & 0 & 0 \\
\end{array}
\eeq
The crucial feature about this answer is that it does not sum to zero when summed over all lattice points. On the other hand, for any TLD operator such a computation would give something which sums up to zero. Hence, $A^{(1)}$ is not a TLD operator.

One may be puzzled that $\calO^{\rm lat}_{\mu}$ is not a TLD operator, while its ``naive continuum limit'' operator given in \reef{eq:Vcand} is a total derivative. In fact there is no contraction. If an operator is TLD, its naive continuum limit will be a total derivative, but the inverse implication does not have to hold. For a very simple example, consider lattice operator
\beq
s(x) s(x+\hat e_1) \nabla_1 s(x)
\eeq
Naive continuum limit $\phi^2\del_1 \phi =\frac 13 \del_1\phi^3$ is a total derivative, but it's easy to check that the lattice operator is not TLD.

 \section{Comments on operator matching}
 \label{sec:matching-rev}
   
   Here we collect some well known facts about operator matching between UV theory and its IR fixed point.
   UV theory may be a lattice spin model, a field theory with cutoff, or a continuum limit field theory.
   
   \subsection{Matching in the lattice spin model}
   
   We consider first the lattice spin model case, and will explain the necessary modifications to UV field theory case later on. For definiteness let us think about the $d=3$ Ising model, on a cubic lattice of spacing $a$ (we could specialize to $a=1$ without loss of generality). We tune the lattice coupling (temperature for the Ising model) to the second-order phase transition. The lattice theory with so finetuned couplings flows, in the RG sense, at large distances to the IR fixed point (IRFP), which we also call ``critical theory". The critical theory has full $O(3)$ invariance, while the lattice theory itself has rotational invariance broken to the cubic subgroup. The critical theory has local operators $\calO_i(x)$ which have well-defined scaling dimensions $\Delta_i$ and transform in $O(3)$ representations. The lattice theory has lattice operators which form multiplets under the lattice symmetry group (cubic group).
   The critical theory is sometimes called CFT, but here we will avoid using this terminology since we don't want to assume conformal symmetry from the start.
The important point is that critical theory correlators are defined at all distances $0<r<\infty$, while correlators of the lattice theory are defined at discrete distances $r\ge a$.
   
How to recover parameters of the critical theory in a lattice simulation? Two issues complicate this extraction. {\bf The first issue} is that operators of the lattice theory, naturally given in terms of lattice variables, do not have well-defined scaling dimension, but should be thought of as linear combinations of such operators. {\bf The second issue} is that the lattice theory, even with couplings finetuned to the second-order phase transition, does not sit precisely at the fixed point, but only flows to it at large distances. Let us consider in turn how these issues manifest themselves.
   
   Consider the simplest lattice operator, spin $S^{\rm lat}(x)$. We should expand it in critical theory operators. The appearing terms will have to be, as $S^{\rm lat}(x)$, $\bZ_2$-odd cubic group singlets. The expansion (sometimes referred to as matching) will have the form:
   \beq
   S^{\rm lat}(x) = A_1\sigma(x)+ A_2 \del^2 \sigma(x)+ A_3 \sigma'(x) +A_4 \del_\mu R_\mu+d_{\mu\nu\lambda\sigma}(A_5
   \del_\mu\del_\nu\del_\lambda\del_\sigma \sigma + A_6 R_{\mu\nu\lambda\sigma})+\ldots
   \label{eq:Sexp}
   \eeq
   There are infinitely many terms but we only wrote the first few representative ones. $\sigma$ and $\sigma'$ are the first two $\bZ_2$-odd scalars of the critical theory (of dimension $\Delta_\sigma\approx 0.518$, $\Delta_\sigma'\approx 5.29$). Derivatives of these operators with indices contracted so that they are scalars can also appear ($\del^2\sigma$ being shown as a representative case). In addition scalar derivatives of tensor $\bZ_2$ operators are also expected to appear, the representative case being the  divergence of some $\bZ_2$ odd vector $R_\mu$ (dimension of the lowest such vector in the critical Ising theory is unknown). All the above terms are $O(3)$ scalars, hence cubic singlets. However, since rotational invariance is broken by the lattice, some tensor operators may appear as long as they are multiplied by tensors which are invariant under the cubic group but not the full $O(3)$. The first such tensor is the rank-4 tensor with nonzero components $d_{1111}=d_{2222}=d_{3333}=1$, and we show two terms involving this tensor, multiplied by $A_{5,6}$.

On a lattice with spacing $a$, all coefficients $A_i$ in this expansion will be given by $A_i=\tilde{A}_i a^{\Delta_i}$, with $\tilde{A}_i$ a dimensionless number and $\Delta_i$  the dimension of the critical operator multiplied by the corresponding coefficient. On a lattice of unit spacing they will be simply $O(1)$ numbers.
   
With the expansion \reef{eq:Sexp}, correlators of $S^{\rm lat}(x)$ in the lattice theory, can be matched with sums of correlators of operators in the critical theory. For example, for the 2pt function we have:
\begin{multline}
\<S^{\rm lat}(x) S^{\rm lat}(y)\>_{\rm lattice}= 
A^2_1\<\sigma(x) \sigma (y)\>+
A_1 A_2\,( \del_x^2 +\del_y^2) \<\sigma(x) \sigma (y)\>
+ A^2_2\, \del_x^2 \del_y^2
\<\sigma(x) \sigma(y)\> \\+ A^2_3\, \<\sigma'(x) \sigma'(y) \> +A^2_4\, \del^x_\mu \del^y_\nu \<R_\mu(x) R_\nu(y)\>+\ldots
\label{eq:SSexp}
\end{multline}
Here the correlator in the l.h.s.~can be measured in a lattice simulation, and by this equation it should be equal to a sum of critical correlators in the r.h.s. Consider for example correlators in infinite volume. The critical theory correlators are expressed in terms of scaling dimensions of the fields. For scalars:
\beq
\<\calO_i(x)  \calO_i(y)\>=\frac 1{|x-y|^{2\Delta_i}}\,,
\eeq
where 1 is just a normalization. For a vector operator we would have
\beq
\<R_\mu(x)  R_\nu(y)\>=\frac {\delta_{\mu\nu}+\alpha (x-y)_\mu (x-y)_\nu/|x-y|^2} {|x-y|^{2\Delta_R}}
\eeq
Here the constant $\alpha$ equals $-2$ in a CFT with $R_\mu$ a vector primary, but in a scale invariant theory but non-conformal theory it could be different. Also in a non-conformal theory there could be nonzero 2pt functions between operators of unequal scaling dimension 
which then have to be added to the r.h.s. of \reef{eq:SSexp}. In any case, according to this discussion, and taking into account the expected size of coefficients $A_i$, the r.h.s. of \reef{eq:SSexp} contains a series of terms decaying with the distance as $const.(a/r)^{p_i}$ where the powers $p_i$ are simply related to scaling dimensions of operators appearing in the r.h.s. of \reef{eq:Sexp}. We see that only dimensionless ratios of distances enter into this expression. If we go to distances $r\gg a$, then the lowest power $p_1=2\Delta_\sigma$ will dominate and the first correction will be suppressed by two more powers of the distance. The terms involving $d_{\mu\nu\lambda\sigma}$ tensor will have nontrivial angular dependence, a sign of rotational symmetry breaking. The leading such term will appear from the crossterm $\<\sigma \del_\mu\del_\nu\del_\lambda\del_\sigma \sigma\>$ and will be tiny, suppressed by 4 powers of the distance. 

To complete the just given discussion, we need to address the above-mentioned {\bf second issue}, taken into account by perturbing the action of the critical theory by irrelevant operators. More precisely, we can describe the system by the action 
\beq
I_{IRFP} + \int d^dx \left[ g_1  \varepsilon'(x)
+ g_2  \varepsilon''(x) +g_3
d_{\mu\nu\lambda\sigma}  L_{\mu\nu\lambda\sigma}(x)+\ldots
\right]
\label{eq:defaction}
\eeq
where  all $\bZ_2$-even irrelevant operators, invariant under the cubic symmetry of the lattice, are present generically. By dimensional analysis, the couplings are given by $g_j =\tilde{g}_j a^{\Delta_j-d}$ where $\tilde{g}_j$ are dimensionless numbers.
The expansion \reef{eq:SSexp} is still true, but correlators in the r.h.s.~should be evaluated in the perturbed theory.
 Specializing again to the 2pt function, the presence of perturbations will lead to the following effect. In addition to the powers $p_i$ occurring in the scale-invariant case there will occur powers $p_i'=p_i+\omega_j$ where $\omega_j=\Delta_j-d$ are all possible correction-to-scaling exponents, with $\Delta_j$ dimensions of irrelevant $\bZ_2$-even operators. The smallest such exponent is $\omega_1 = \Delta_{\vareps'}-3\approx 0.83$.\footnote{The idea of improved lattice actions is to use models that allow to tune to zero the couplings of the first few leading irrelevant operators in \eqref{eq:defaction}. For example, the Blume-Capel model used in \cite{Hasenbusch0} allows to set $g_1=0$ thus removing the leading corrections to scaling due to $\vareps'$.} Some of these power law corrections will come with nontrivial angular dependence. This is to be expected, since the lattice theory breaks rotation invariance. The smallest rotational invariance breaking exponent $\omega_{3}\approx 2.02$ is related to the dimension of the lowest $\bZ_2$-even cubic group singlet that is not an $O(3)$ scalar. In the case of 3d Ising, this is the lowest $\bZ_2$-even spin-4 operator $L_{\mu\nu\lambda\sigma}$ contracted with the $d_{\mu\nu\lambda\sigma}$ tensor (while $R_{\mu\nu\lambda\sigma}$ in \reef{eq:Sexp} was $\bZ_2$-odd).
 
Matching can also be done for lattice operators transforming in nontrivial representations of the lattice symmetry group, vector being our main case of interest. The $(d=3)$-dimensional vector representation is irreducible both under $O(3)$ and under the cubic group. For a generic $\bZ_2$-even lattice vector operator $\calO^{\rm lat}_\mu$, some representative terms in its expansion will be:\footnote{The coefficients $A_i$ are of course not the same as for $S^{\rm lat}(x)$.}
\beq 
\calO^{\rm lat}_\mu = A_1 V_\mu + A_2\del_\mu \vareps + A_3 \del_\mu T'_{\mu\nu} +A_4 
d_{\mu\nu\lambda\sigma} R_{\nu\lambda\sigma}+\ldots
\eeq
This indicates that the r.h.s.~can contain vector critical operators $(V_\mu)$, derivatives of scalars $(\del_\mu\eps)$ and divergences of tensors ($\del_\mu T'_{\mu\nu}$, excluding the stress tensor $T_{\mu\nu}$ as it is conserved), as well as rotation-invariance breaking terms involving higher-rank tensors contracted with special  tensors like $d_{\mu\nu\lambda\sigma}$, to get objects which transform correctly under the cubic group.

In the generic case we expect all $A_i=O(a^{\Delta_i})$ as for $S^{\rm lat}(x)$. In the special case of $\calO^{\rm lat}_\mu(x)$ being a total lattice derivative,\footnote{Total lattice derivatives are local operators that when summed over a  region of the lattice, reduce to operators at the boundary of that region.
For example,  $\nabla_\mu \phi(x)=[\phi(x+a e_\mu)-\phi(x-a e_\mu)]/(2a)$ is a lattice derivative. See Appendix \ref{total}.}  we will have $A_1=A_4=0$ and only the terms like $A_{2},A_3$ could contribute.

We emphasize that all we know of the coefficients $A_i$ on general grounds is that they are $O(a^{\Delta_i})$ numbers. There is no simple theoretical way to determine these numbers apart from a lattice simulation. All operators which are allowed by lattice and internal symmetries (and total lattice derivative constraints) will appear in the r.h.s. The problem of determining these coefficients is a ``long distance" problem: it has to do with how the microscopic theory approaches the IR fixed point at long distances. 

  \subsection{Matching in the lattice field theory}

It is instructive to consider what changes when we replace the spin model by the latticized $\phi^4$ field theory, defined by the lattice action
\beq
a^3 \sum_x \left[\frac 12 \sum_{\mu =1}^3 (\nabla_\mu \phi(x))^2+ m^2 \phi(x)^2 +\lambda \phi(x)^4 \right]\,.
\label{eq:LFT}
\eeq
where $\nabla_\mu \phi(x)=[\phi(x+a e_\mu)-\phi(x-a e_\mu)]/(2a)$ is the lattice derivative.
For each value of the quartic coupling $\lambda>0$ we can find a value of the mass parameter corresponding to a second-order phase transition. For this value $m^2_*(\lambda)$ the theory flows at large distances to the critical theory, which does not depend on $\lambda$ and is actually the same as for the Ising spin model. The operators of the UV theory can be then expanded in critical theory operators. For example, we can write an expansion for $\phi(x)$ of the same form \reef{eq:Sexp} as for the spin operator $S^{\rm lat}(x)$. The symmetry reasoning which led to this expansion remains the same, and the same operators will appear in the r.h.s. However, the discussion of the size of coefficients $A_i$ has to be slightly modified.

We say that the $\phi^4$ theory is strongly coupled at the lattice scale if the quartic coupling $\lambda$ is not small. The appropriate dimensionless condition in 3d is $\lambda a \gtrsim 1$.\footnote{Notice that the lattice field $\phi(x)$ has dimension $1/2$ like a free scalar field in 3d. This implies that the quartic coupling $\lambda$  has mass dimension 1.} The effects of such largish quartic coupling are strongly felt already at the lattice scale (and a fortiori at all longer distance scales). Because of this, the RG flow will converge to the IR fixed point at distances $r$ not much higher than $a$. The matching coefficients in the strongly coupled latticized $\phi^4$ theory will thus be of the same generic size as for the spin Ising model, \emph{i.e.} $A_i =O(a^{\Delta_i})$. 

\begin{figure}[htb]
	\centering
	\includegraphics[width=0.6\textwidth]{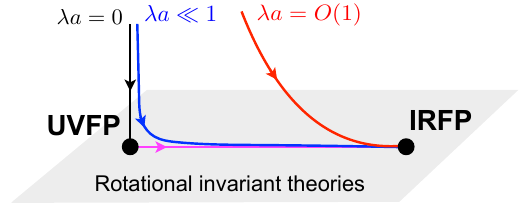}
	\caption{Various RG flows on the critical surface of the latticized $\phi^4$ field theory. All flows with $\lambda>0$ end up in the IRFP because we tuned the mass to its critical value. However, flows that start with $\lambda a \ll 1 $ will first be attracted to the UVFP and from there move to the IRFP. More precisely, if the quartic coupling is parametrically small at the UV scale $a$, the RG flow will be controlled by the UVFP until the scale $\ell_0=1/\lambda$. At this scale, the flow transitions to the neighborhood of the IRFP. The flow with $\lambda=0$ corresponds to a quadratic theory which ends in the UVFP once the rotation invariance breaking terms have decayed. 
	}
	\label{figflow}
\end{figure}

If on the other hand the quartic satisfies $\lambda a \ll 1$, the starting point of RG flow finds itself not far from the gaussian UV fixed point (UVFP). The RG trajectory can then be divided into two parts (see Fig.~\ref{figflow}). 
In this case we say that the UV lattice theory is `weakly coupled'. The first part of the RG flow happens in the neighbourhood of the UVFP. It corresponds to distances $\ell \ll \ell_0$, where $\ell_0 = 1/\lambda \gg a$. The second part starts at distances $\ell\sim \ell_0$ where the flow transitions from the neighbourhood of free UVFP to the strongly interacting IRFP.

In the first part of the flow we can approximate the action of the flowing theory expanding around the UVFP action in perturbations parametrized by normalized operators of the gaussian theory:
\beq
I= I_{UVFP}+ \int d^3x\,\left[ 
u_1 :\phi^2(x): +
u_2 :\phi^4(x): 
+u_3  d_{\mu\nu\rho\sigma} :\phi \partial_\mu \partial_\nu \partial_\rho \partial_\sigma \phi(x) : 
+\dots \right]
\label{eq:expGauss}
\eeq
where $u_i= \tilde{u}_i a^{\Delta_i^{UVFP}-3} $ with $\tilde{u}_i $ dimensionless.
Generically, we expect all $\tilde{u}_i =O(1)$. However, for weakly coupled flows we have $\tilde{u}_2 \sim \lambda a \sim a/\ell_0 \ll1$. Furthermore, because we tuned the mass term to its critical value we also have $\tilde{u}_1 
\ll1$. The first term breaking rotational invariance has $u_3=\tilde{u}_3 a^2$ with $\tilde{u}_3=O(1)$.\footnote{The couplings of irrelevant operators that involve more than two powers of $\phi$ are also suppressed by the small parameter $\lambda a$ because at $\lambda=0$ the lattice path integral is exactly gaussian.}

The second part of the flow starts at the scale $\ell_0 = 1/\lambda \gg a$.
Therefore, the scale $\ell_0$ plays the role of UV cutoff for the second part of the flow. It is then useful to write $u_i 
=\bar{u}_i \ell_0^{\Delta_i^{UVFP}-3}$ to define dimensionless couplings $ \bar{u}_i$ with respect to the UV cutoff for the second part of the flow. 
This gives $ \bar{u}_2 = O(1)$ for the quartic coupling and $\bar{u}_3 \sim (a/\ell_0)^2 \ll 1$ for the leading irrelevant coupling that breaks rotational symmetry.
The second part of the flow can then be described using the action \eqref{eq:defaction}  with dimensionless couplings $\tilde{g}_i$ defined by $g_i = \tilde{g}_i \ell_0^{\Delta_i^{IRFP} -3}$. We expect $\tilde{g}_1 \sim \tilde{g}_2 \sim O(1)$ and $\tilde{g}_3 \sim \bar{u}_3\sim (a/\ell_0)^2 \ll 1$.

We thus see that the second part of RG flow starts with some irrelevant operators in the action having dimensionless couplings much smaller than the other ones. This effect was absent in the spin lattice model case, where all irrelevant operators were expected to be present at the cutoff scale with $O(1)$ coefficients  in lattice units. As a consequence, rotation breaking in the IR, already small in the spin model case, will be even further suppressed in the weakly coupled lattice field theory case.

Now let us discuss matching of operators, which also happens in two stages. First we expand lattice field theory operators into operators of the UVFP. E.g. we will have
\beq
\phi^{\rm lat}=A_1 \phi + A_2 :\phi^3:+\ldots
\eeq
Coefficients of this expansion have a power series expansion in $\lambda$. For example we expect $A_1=1+O(\lambda a)$, while $A_2=O(\lambda a^2)$. Then we have to expand UVFP operators in IRFP operators. This matching is done at the scale $\ell_0$. E.g. we have:
\beq
(\ell_0)^{\Delta_\phi} \phi = B_1 (\ell_0)^{\Delta_\sigma} \sigma +B_2 (\ell_0)^{\Delta_{\sigma'}} \sigma'+\ldots
\eeq
Since this matching is done at the scale where the flow is strongly coupled, the coefficients $B_i$ cannot be easily predicted and are expected to be $O(1)$. Combining the two matchings, we will get expressions for lattice field theory operators in terms of IRFP operators.

\section{Possible boundary conditions}
\label{boundary}

One can imagine modifying our setup described in the main text, by changing the boundary conditions at $x_3=0,L-1$. The purpose would be to find boundary conditions which lead to an even larger $f(t)$ and thus improve the signal-to-noise ratio. It makes sense to keep translation invariance in the $x_2$ direction, so that $\langle I(x_2,x_3)\rangle$ is $x_2$ independent and can be averaged in this direction.

As discussed in the main text, we have to break the $x_1$ flip symmetry. One way to do this is to choose different boundary conditions for different parts of the $x_3=0,L-1$ boundaries, depending on $x_1$.

In addition to the free and fixed boundary conditions (b.c.) described in the main text, there are two other imaginable types of b.c.~worth discussing. 

\subsection{Gluing b.c.} 

The gluing b.c.~changes topology of our manifold, by gluing one part of the boundary to another.
For example, one can imagine gluing the gray parts of the $x_3=0,L-1$ boundaries in Fig.~\ref{bc}, instead of imposing the fixed b.c.~there. In practice, gluing is achieved by identifying points pairwise or, equivalently in the large $L$ limit, by creating links joining the points being glued. In the just mentioned example, we would be identifying points 
\eqn
{
(x_1,x_2,x_3=L-1)\quad\text{with}\quad (x_1+L/4,x_2,x_3=0)\quad(0\le x_1< L/2,0\le x_2<L)
}
Gluing does not have to preserve order, for example we could have instead chosen to glue the gray parts of the boundaries while simultaneously flipping the $x_1$ coordinate.
Such a reversed gluing would be a different boundary condition.

One can even glue parts of the same boundary, e.g.~the lower and upper white parts of the $x_3=0$ boundary in Fig.~\ref{bc} (again, in the direct or the reversed $x_1$ order).  

\subsection{Changing the strength of boundary interactions}
We may change the strength of interaction among spins belonging to some part of the boundary to $\beta_{\rm bdry}\ne \beta_c$. Two particularly interesting values of $\beta_{\rm bdry}$ are as follows.
\begin{itemize}
\item
$\beta_{\rm bdry}=\beta_{\rm sp}\approx 0.33302$. This fixes $\beta_{\rm bdry}$ to the value corresponding to the ``special" boundary phase transition. Recall that the special transition separates the ``ordinary" boundary behavior for which the boundary remains disordered at the critical temperature, from the ``extraordinary" one when the boundary is ordered at the critical temperature. The ordinary (extraordinary) behavior is realized at $\beta_{\rm  bdry}<\beta_{\rm  sp}$ ($\beta_{\rm  bdry}>\beta_{\rm  sp}$). The $\beta_{\rm  sp}$ for the 3d Ising model given above was determined in \cite{Hasenbusch-sp}. Since the boundary points have fewer neighbors than the bulk points, $\beta_{\rm bdry}=\beta_c$ belongs to the ``ordinary" phase, and this explains why $\beta_{\rm sp}> \beta_c$. 
\item
$\beta_{\rm bdry}=\infty$. This enforces that all spins are equal along a part of the boundary, which is the maximally efficient way to enforce the ``extraordinary" boundary behavior. Notice that unlike the fixed boundary condition, the spins can still fluctuate between $\pm1$, but only all at once. This difference may seem minor, but it has the following practical consequence. The fixed b.c.~can be used if the simulations are performed using the Metropolis algorithm, as in the main text. On the other hand, if the simulations are performed using cluster algorithms, it leads to lowering the acceptance rate since clusters which touch the boundary cannot be flipped. The $\beta_{\rm bdry}=\infty$ boundary condition does not have this difficulty.
\end{itemize}
There are many imaginable combinations of the four boundary condition types which break symmetries of the lattice in a way which makes $f(t)$ nonzero. It is tedious to simulate one by one all possible combinations for the Ising model and see which one gives the largest $f(t)$. It would be nice to have a way to guess a good boundary condition. A heuristic method is described in the next appendix.

\section{Heuristic optimization of boundary conditions}
\label{heuristic}

Consider the free massless scalar theory on the cubic lattice, described by the action:
\eqn{
H=\sum_{\langle x y \rangle} (\phi(x)-\phi(y))^2\,,\quad \phi(x)\in\bR\,.
}
We consider in this theory a lattice operator $V_{\mu}^{\rm lat}$ given by the same equation \Vlat with $\phi(x)$ instead of $s(x)$. We make a heuristic hypothesis that one can get an idea about the size of $\langle I\rangle$ in the critical Ising model by measuring the same quantity in the free scalar theory on the same cubic lattice. One motivation for this hypothesis is that in $d=4$ the two theories are actually identical. We won't attempt to justify this hypothesis any further. It's amusing that empirically it seems to work. Once the b.c.~is so heuristically guessed, the actual hard computation will be an honest Monte Carlo simulation in the 3d Ising.

To use the heuristic, we have to establish a correspondence between boundary conditions for the two models. This correspondence is as follows:
\begin{enumerate}
\item
The free b.c.~in the Ising corresponds to the Dirichlet b.c.~for the free scalar. Indeed, the free b.c.~in Ising leads to the ``ordinary" boundary behavior, where the order parameter is effectively zero on the boundary \cite{Cardy:1996xt}.

\item
The gluing b.c.~in Ising clearly corresponds to the same gluing for the free scalar.

\item 
$\beta_{\rm bdry}=\infty$ for the Ising corresponds to imposing that $\phi(x)$ remains constant on this part of the boundary for the scalar.

\item 
$\beta_{\rm bdry}=\beta_{\rm sp}$ for the Ising corresponds to the Neumann (i.e.~free) boundary condition for the scalar \cite{Cardy:1996xt}.

\item
The fixed 3d Ising boundary condition can be modeled by adding a constant magnetic field (linear in $\phi(x)$ term) on the boundary, pushing the free scalar in the needed direction.
\end{enumerate}

We won't give full details on how one actually performs the calculation for the free scalar. This calculation is inexpensive since one is computing a gaussian path integral. One constructs the lattice action, evaluates the Green's function, and finally evaluates the observable. The computation is done numerically and takes only a few seconds for a given boundary condition. The most expensive step is the Green's function evaluation which requires to invert an $L^3 \times L^3$ matrix. 

After playing with the free scalar, we concluded that the boundary condition in Fig.~\ref{bc} is particularly promising. Notice that since we have the same fixed b.c.~on two parts of the boundary, and since we measure a $\bZ_2$-even observable, for the purpose of the heuristic computation we could replace the fixed boundary condition with $\beta_{\rm bdry}=\infty$. 

Before we discovered the heuristic optimization trick, we tried other boundary conditions in the 3d Ising, but they led to a smaller $f(t)$. 

We could have just postulated the boundary condition in Fig.~\ref{bc}, but we prefer to play in the open. This is because we have not performed exhaustive optimization. Even better b.c.~likely exist, and our heuristic may be helpful to search for them. 




\end{document}